\documentclass[twocolumn]{article} 

\usepackage{graphicx}
\usepackage{pdfpages}

\usepackage[utf8]{inputenc}
\usepackage[a4paper,width=170mm,top=25mm,bottom=25mm]{geometry} 
\usepackage{fancyhdr}
\renewenvironment{abstract}
{\centerline{\large\bfseries Abstract}\vspace{0.5ex}\bgroup\leftskip 20pt\rightskip 20pt\small\noindent\ignorespaces}
{\par\egroup\vskip 0.25ex}

\usepackage{bbm}

\usepackage{fontawesome}
\usepackage[edges]{forest}

\usepackage{subfig}

\usepackage{amsmath}

\usepackage{ amssymb }

\usepackage[english]{babel}
\usepackage{amsthm}

\usepackage{graphicx} 
\usepackage[skip=5pt]{parskip} 

\usepackage{tikz}
\usetikzlibrary{shapes.geometric, arrows}
\usetikzlibrary{positioning}

\usepackage{mathtools}

\usepackage[sorting=none]{biblatex}
\usepackage{titlesec}
\titleformat{\chapter}[display]   
{\normalfont\huge\bfseries}{\chaptertitlename\ \thechapter}{20pt}{\Huge}   
\titlespacing*{\chapter}{0pt}{-50pt}{0pt}

\title{Digital Histopathology with Graph Neural Networks: Concepts and Explanations for Clinicians} 
\author{Alessandro Farace di Villaforesta, Lucie Charlotte Magister, Pietro Barbiero, Pietro Li\`{o}}
\date{October 2023}

\begin{document}

\maketitle
\begin{abstract}
     Addressing the challenge of the ``black-box" nature of deep learning in medical settings, we combine GCExplainer - an automated concept discovery solution - along with Logic Explained Networks to provide global explanations for Graph Neural Networks. We demonstrate this using a generally applicable graph construction and classification pipeline, involving panoptic segmentation with HoVer-Net and cancer prediction with Graph Convolution Networks. By training on H\&E slides of breast cancer, we show promising results in offering explainable and trustworthy AI tools for clinicians.
\end{abstract}
\section{Introduction}





%

Cancer is a leading cause of death worldwide, accounting for nearly 10 million deaths globally in 2020 \cite{who:cancer}. Early diagnosis and intervention plays an important role in both improving patient outcomes and reducing treatment costs \cite{miller2012screenings}. 

Recent work has shown the effectiveness of analysing signalling interactions between tumour cells and their surrounding microenvironment (TME) in understanding tumour progression and metastisis \cite{jahanban2018modulating}. Early automated diagnostic systems applied computer vision techniques to Hematoxylin and Eosin (H\&E) stained slides, but more recent work has proposed incorporating spatial and structural information explicitly into learning algorithms \cite{wang2021cell}; this can be done by extracting cell or tissue graphs which are ammenable to graph neural networks.


Despite all of their advances, Artificial Intelligence models fail to gather significant adoption by medical practitioners. Some \cite{kundu2021ai} have pointed to the ``black-box'' nature of these models as the limiting factor; models are fed an input and, as if by some inscrutable magic, come to a decision. When placed in a high stakes environment like a hospital, doctors want such models to instil confidence not doubt. It is for this reason that the work in explainable AI \textbf{(XAI)} is incredibly topical.


GNNExplainer \cite{ying2019gnnexplainer}, a prominent example of post-hoc explainbility, produces local or instance-level explanations to justify model predictions. This means explanations are specific to that particular example and do not originate from some prior understanding of the task; it is left to the user to draw meaningful connections, and this is susceptible to confirmation bias \cite{adebayo2018sanity}. A recent taxonomy \cite{yuan2022explainability} illustrates the sparsity of work on model-level explanations over the instance-level explanations, and we point to GCExplainer \cite{magister2021gcexplainer} as one of the only methods of discovering global concepts.


\textbf{Contributions}: Our principle contribution was to fuse GCExplainer with Logic Explained Networks \cite{ciravegna2021logic} to provide reasoned model-level explanations, and apply it to a medical dataset.


\section{Methodology}

Our approach is made up of three components that can be independently developed and optimized: graph construction, graph classification and post-hoc explanation. 

\subsection{Graph Construction}
The principal requirement to synthesise cell graphs is to identify the cells/nuclei present in an H\&E stained slide. The most difficult aspect with instance segmentation of nuclei is separating neighbouring or overlapping nuclei; this problem is broadly referred to as \textbf{under-segmentation} where too few segments have been defined. To overcome this, as well as classify such cells, we employed HoVer-Net \cite{grahamHoVerNetSimultaneousSegmentation2019}, a panoptic segmentation architecture. Their key contribution over previous encoder-decoder architectures was to jointly train three dedicated branches:

\begin{itemize}
    \item \textbf{Nuclear Pixel (NP) branch}: Generates a cell mask by performing semantic binary segmentation.
    \item \textbf{HoVer branch}: Learns the displacement of pixels from their nucleus' centroid. The head of the decoder has two channels, one for learning horizontal displacements and one for vertical. These values range from -1 for far left/top to 1 for far right/bottom. The outputs are collectively referred to as the two \textbf{hover maps}.
    \item \textbf{Nuclear Classification (NC) branch}: Generates a semantic multiclass segmentation for predicting the cell type.
\end{itemize}

In order to create an instance segmentation, the hover maps are translated to energy landscapes using sobel filters and the watershedding algorithm is applied in conjunction with the cell masks to separate touching nuclei. Finally, a majority vote is conducted at each cell to derive its classification.

Next, we define our node embeddings. We took a 64x64 pixel crop of each cell and extracted a series of morphological and biological features. Firstly, we include the Gray-level Cooccurrence Matrix (GLCM) \cite{haralick1973textural} used for capturing textural information; it is particularly effective in analysing, for example, the surrounding stroma tissue which undergoes desmoplasia as a reaction to cancerous cells. We also trained a ResNet-50 \cite{he2016deep} autoencoder to generate a low dimensional representation of the cell image. Lastly, we one-hot encoded the cell classification, and concatenated these features to engineer our embedding.

\begin{figure}[h]
    \centering
    \includegraphics[height=12cm,keepaspectratio]{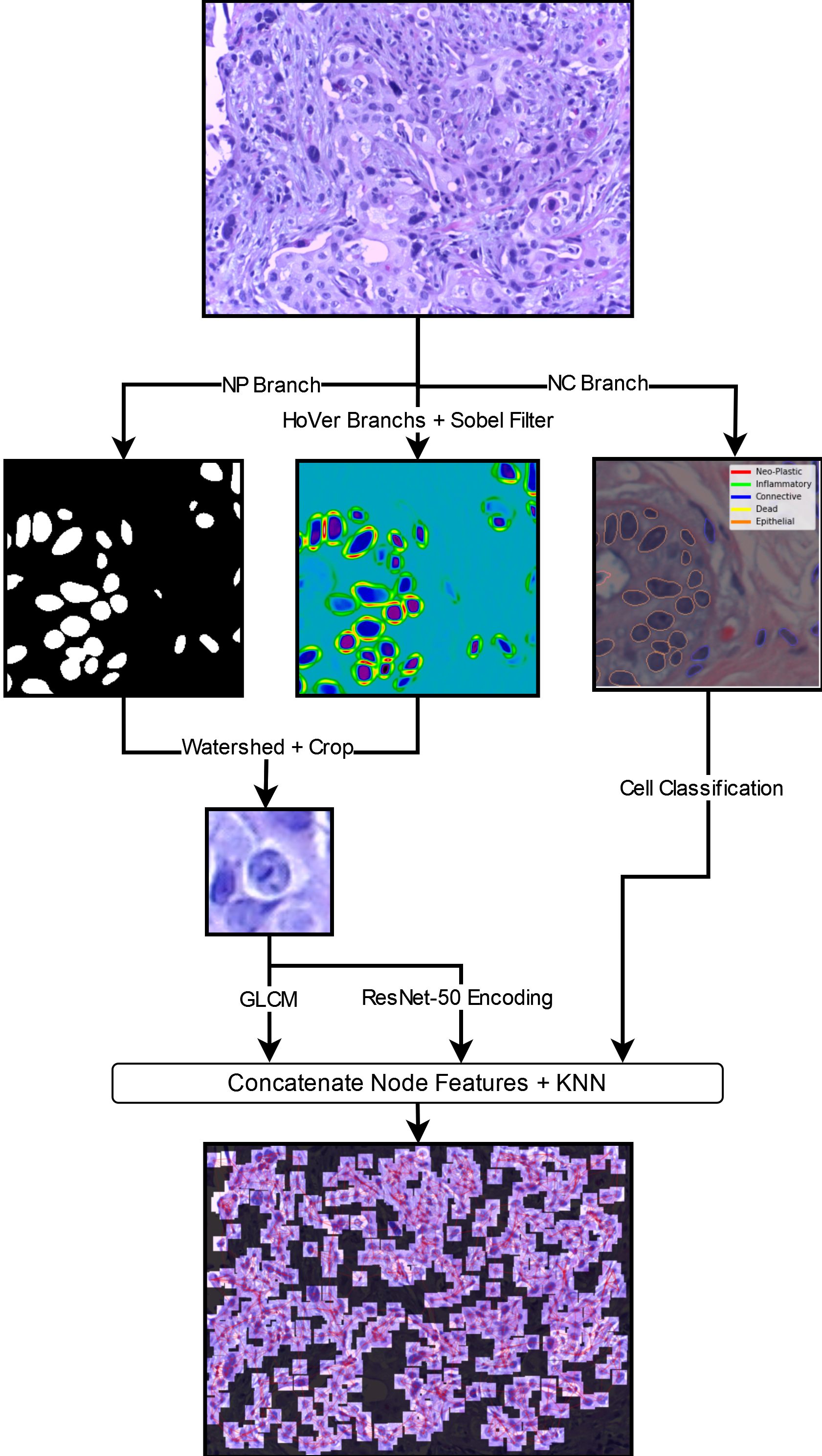}
    \caption{\textit{Graph construction pipeline summarizing the cell graph synthesis from an H\&E stained slide image. Cell nuclei are segmented and classified using the HoVer-Net decoder branches, along with watershedding. Cell crops are taken and morphological features extracted. These embeddings are combined with a KNN graph to create the cell graph. Image `iv013` from the BACH dataset \cite{aresta2019bach} is shown here and used as our running example. }}
\label{fig:graph_classification_pipeline}

    \label{fig:graph_construction}
\end{figure}


\subsection{Graph Classification}

In search of general applicability across a gamut of diseases, we used a Graph Convolution Network \cite{kipf2016semi} as our principle architecture. The GCN layer is particularly effective when applied to homophilic graphs which, because of the biologicaly signalling between neighbouring cells, we hypothesise exists to an extent in our cell graphs. The latent embedding of node $v$ in layer $l+1$ is given by:
\[
h_v^{(l+1)} = \sigma\left( W^{(l)} \sum_{u \in \mathcal{N}(v) \cup \{ v \}} \frac{h_u^{(l)}}{\sqrt{\text{deg}(u) \text{deg}(v)}} \right)
\]
Where $\mathcal{N}(v)$ is the set of neighbouring nodes of $v$, and $W^{(l)}$ is the layer $l$'s trainable weight matrix. After passing the graph through several convolution layers, a global mean pool is taken and then fed into a Multi-Layer Perceptron that will predict if cancer is present.
 

%

\subsection{Post-Hoc Explanation}

\subsubsection{GCExplainer}
Unlike GNNExplainer, GCExplainer\cite{magister2021gcexplainer} generates global, model-level explanations for GNNs by discovering and recognising concepts, which are high-level units of information used to reason about the data. In general, concepts may be extracted in a supervised manner, requiring explicitly defining concept set, or unsupervised. GCExplainer, like its inspiration \cite{ghorbani2019towards}, uses clustering on the activation space to automatically discover concepts.

In order to discover concepts implicitly inferred by the model, GCExplainer suggests taking the activation space of the last neighbourhood aggregation layer in the GNN and performing k-Means clustering on it; it is argued that the nodes within the same cluster are examples of the same deduced concept. This mapping of each node’s latent representation to its nearest cluster effectively translates into learning a global function $\hat{f}:X\longrightarrow C$, where $X$ represents the space of input nodes and $C$ the concept space. To exemplify a particular concept $c$, we take an n-hop subgraph centred around cells whose activations are close to $c$'s centroid.

Since we are performing graph classification, we need to aggregate concepts across the whole graph to create a representative graph-level concept vector. We achieved this by summing-up the one-hot encoding of all present concepts, and then dividing by the count of the most prevalent.

\subsubsection{Logic Explained Network}
\begin{figure*}[t]
\centering

    \includegraphics[width=0.7\textwidth,keepaspectratio]{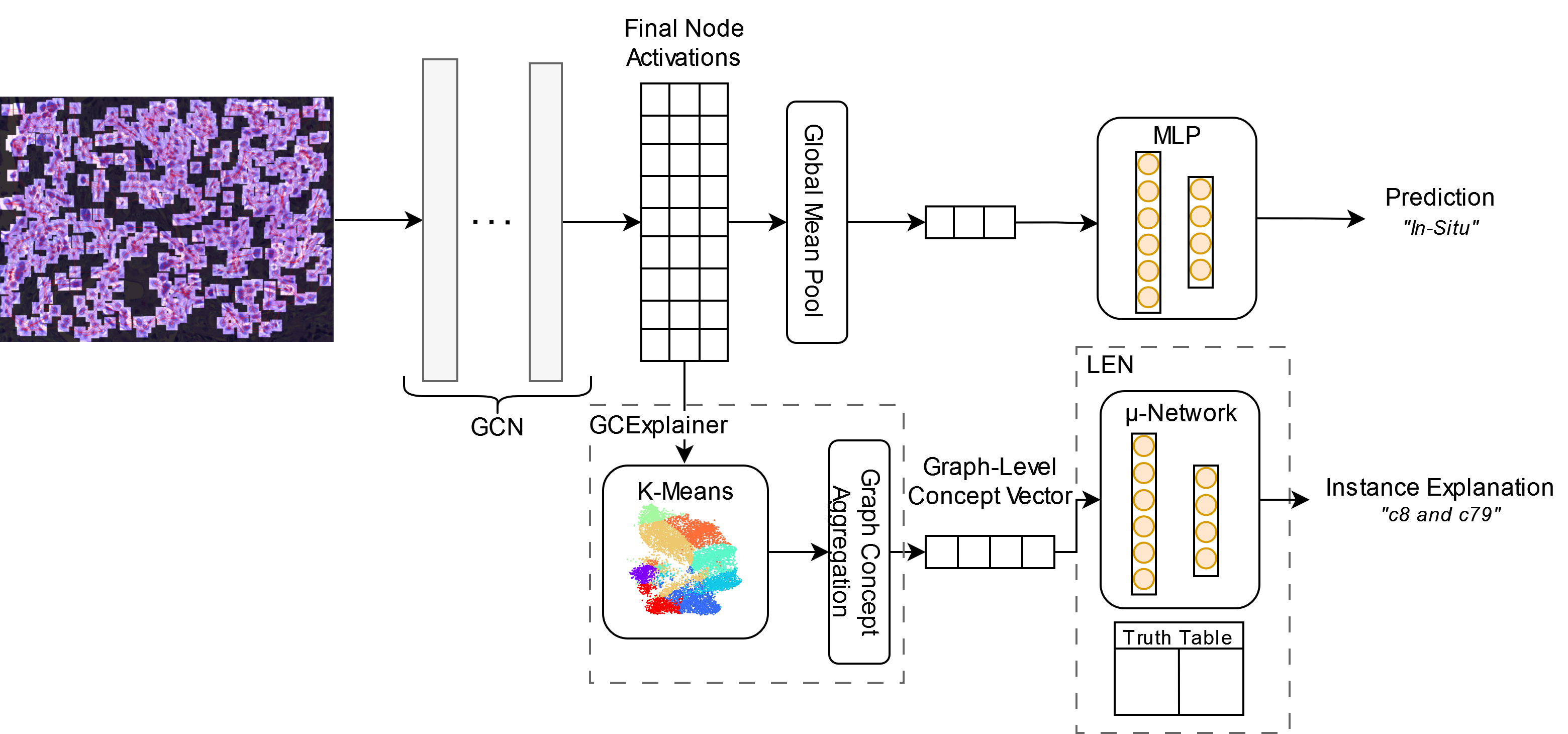}%

\caption{\textit{The graph classification and explanation pipeline. The synthesised graphs are passed through several GCN layers, returning latent node embeddings. This is mean-pooled and the resulting vector is passed through an MLP classifier. For explaining predictions, the activations are assigned to the closest k-Means clusters (found from the training set). After aggregating, the trained LEN provides a local-explanation based on the presence or absence of concepts in the image.}}
\label{fig:explainability_pipeline}  
\end{figure*}

Logic Explained Networks (LENs) \cite{ciravegna2021logic} are neurosymbolic architecture that provide reasoning for classifications in terms of First-Order Logic (FOL) formulas consisting of input concept predicates.

When a neural network implements a mapping \( f: [0,1]^d \to [0,1]^r \), it inherently acquires a logical interpretation. A network satisfying this requirement can be a candidate for a LEN: a feed forward network with suitable activation functions suffices. The logical interpretation can be elucidated by inputting booleanized vectors into the model, which results in the generation of a truth table. This table effectively encodes the network's behavior, leading to the definition of $r$ corresponding logical functions \( \bar{f}_y: \{0,1\}^d \to \{0,1\} \), one for predicting each output class.

The local or example-level explanation for an input is given as the conjunction of the booleanized inputs: 

\[\phi_{\text{local}}(c) = \bigwedge_i \tilde{c}_i \quad \text{where } \tilde{c}_i = \begin{cases} 
\bar{c}_i, & \text{if } c_i > 0.5 \\
\neg \bar{c}_i, & \text{if } c_i \leq 0.5 
\end{cases}\]

where \(\bar{c}_i\) represents the logical value of concept \(i\) from the input concept vector \(c\). The global or class-level explanation is given as the disjunction of all local-explanations for that class, namely:

\[\phi_y = \bigvee_{\bar{c} \in \bar{f}_y^{-1}(1)} \phi_{\text{local}}(\bar{c})\]

Where \( \bar{f}_y^{-1}(1) \) is the set of all logical vectors that lead to class \( y \) being predicted, as recorded by the truth table.


Another desirable quality of explanations is parsimony. To articulate how LENs achieve this, we introduce here the `$\mu$ network', an out-of-the-box MLP-based LEN which strikes a good balance between predictive power and interpretability: we also use this network as our LEN definition. Parsimony is achieved in the $\mu$ network by: $(i)$ using strong lasso regularization \cite{tibshirani1996regression} on the first layer's weights to perform a kind of concept-selection $(ii)$ pruning all nodes whose departing weights' L2-norm is below a relative threshold.


\section{Experiments}
We made use of multiple datasets when training our models. Firstly, we ingested the \textbf{PanNuke} dataset \cite{gamperPanNukeDatasetExtension2020} to train HoVer-Net as it contains segmentation masks for nearly 200,000 labelled nuclei across over 7,000 images. Samples range across 19 different tissue types and 5 cell types: neoplastic, epithelial, inflammatory, connective and dead. To train our classifer, we used graphs derived from the BreAst Cancer Histology images (BACH) \cite{aresta2019bach} dataset. It consists of 400 H\&E images of breast tissue, with 100 examples for each of the four classes (with each class containing 100 examples): Normal, Benign, In-Situ, and Invasive.

After taking a 80/20 split and removing any examples where a cell graph could not be synthesised, we were left with 310 training and 78 test instances. To ameliorate overfitting, we employed several data augmentation techniques including: node embedding perturbation that adds suitably scaled noise vectors to the input, node dropout, and random position jitter to the nodes. The subsequent set of nodes is passed into a k-Nearest Neighbours to procure the graph. 

\begin{figure}[h]
    \centering
    \includegraphics[width=0.2\textwidth,keepaspectratio]{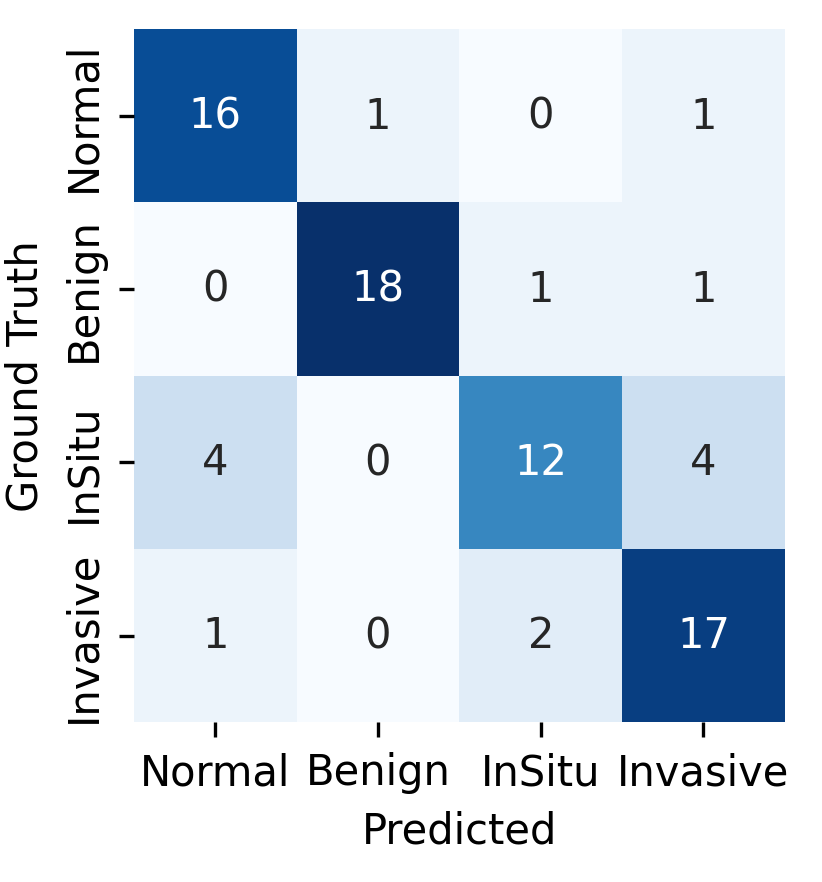}
    \caption{GCN confusion matrix on validation set}
    \label{fig:confusion_matrix}
\end{figure}

We achieved a 80.7\% 4-class accuracy and 89.7\% 2-class accuracy (cancer vs not-cancer) which was sufficient for the purpose of this work.




\section{Discussion}
We found that the combination of GCExplainer with LEN had the capacity to extract meaningful concepts. To illustrate this using our running example, the LEN noted that `iv013' is invasive because of the strong presence of concepts 8 \& 79 when trained with 128 concepts. After pinpointing the nodes whose activations fall closest to these concepts, we compared these to exemplary 3-hop subgraphs discovered during training which we have mapped back onto the original image.

For concept 8, we observe these central cells are all likely undergoing mitosis. For concept 79, we note a particularly complex and heterogeneous nucleus microstructure present, with multiple nucleolus as well as potentially vesicles. 

These were not the only concepts extracted from the graph: there were 17 other concepts present, some with higher pervasiveness than the two concepts returned. However, the LEN had not considered these concepts as contributing as much to the decision process and so had been pruned out. 

\begin{figure}[h]
    \centering
    \includegraphics[width=0.4\textwidth,keepaspectratio]{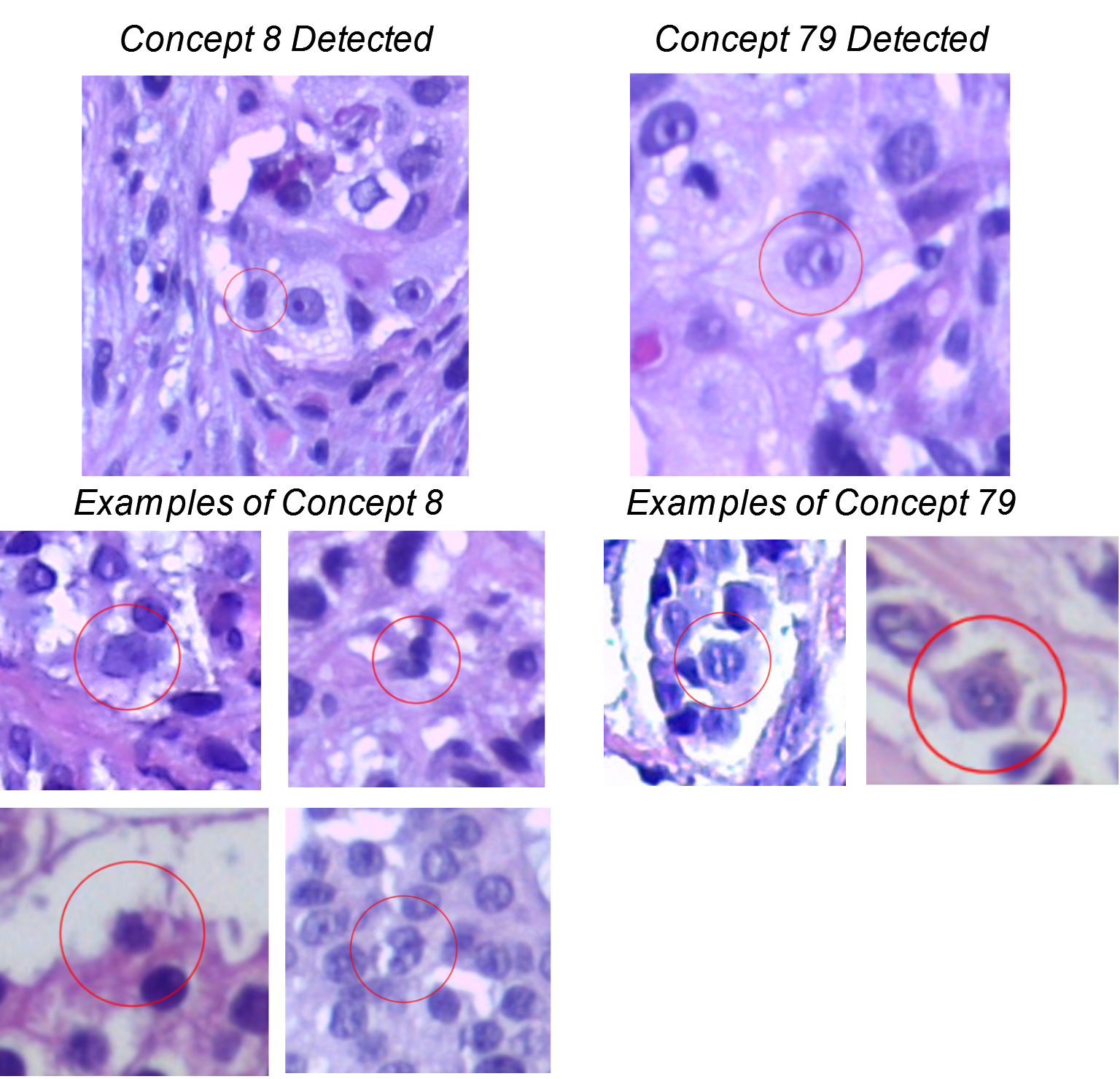}
    \caption{\textit{Visualizations of concepts 8 \& 79. The crops at the top are from iv013, while the crops from the bottom are from the training set; for clarity, we have circled and zoomed into the centre node of the representative subgraphs.}}
    \label{fig:concepts8-79}
\end{figure}

\begin{table}[h]
\centering
\setlength{\tabcolsep}{5pt} 
\renewcommand{\arraystretch}{1.2} 
\begin{tabular}{|c|c|c|c|}
\hline
k \textbackslash $\lambda$ & 1 & 0.1 & 0.01  \\
\hline
4 & $72.76 \pm 0.69$ & $71.03 \pm 2.76$ & $73.79 \pm 2.53$\\
\hline
16 & $74.48 \pm 1.29$ & $74.48 \pm 1.29$ & $77.93 \pm 2.29$ \\
\hline
32 & $79.31 \pm 1.09$ & $78.28 \pm 0.84$ & $70.00 \pm 3.20$\\
\hline
64 & $78.28 \pm 1.76$ & $78.62 \pm 1.38$ & $70.00 \pm 2.80$\\
\hline
128 & $78.62 \pm 1.38$ & $80.34 \pm 0.84$ & $67.93 \pm 1.38$\\
\hline
256 & $82.07 \pm 1.76$ & $82.41 \pm 0.69$ & $71.38 \pm 0.84$0\\
\hline
512 & $74.48 \pm 1.29$ & $73.45 \pm 0.84$ & $61.72 \pm 1.29$\\
\hline
\end{tabular}
\caption{Mean ± Standard Deviation completeness scores for different number of concepts $k$ and $L_1$ weights $\lambda$}
\label{table:data_compact}
\end{table}


These explainability methodologies afford engineers, in collaboration with clinicians, the ability to fine tune parameters. On the one hand, a number of concepts $k$ can lead to higher concept completeness scores \cite{yeh2019concept}, which is a measure for how predictive power using the concepts alone. This comes at the cost of: $(i)$ decreasing quality of clustering as observed by a reduction in average silhouette scores \cite{rousseeuw1987silhouettes} from $0.355$ at $k=4$ to $0.191$ at $h=512$, $(ii)$ a more complicated and longer instance-level explanations due more min-terms. We found that increasing the degree of $L_1$ regularization assisted in both improving accuracy as well as reducing explanation complexity. 








\section{Conclusion} 
We recognise there must be common ground between the language of clinicians and the models to improve adoption of deep learning prediction systems. To this end, the next steps after automated concept discovery is the assignment of textual interpretations by experts. Through a `Doctor-in-the-loop', multiple components of prediction can be vetted, from the purity of concept meanings, to the detection of present concepts, and also to the significance attributed to these concepts.

In regards to future work, we believe the quality of clustering plays a pivotal role in automated concept based discovery. The Concept Encoder Module \cite{magister2022encoding} - which similarly incorporates LENs - can potentially rectify this by `explaining-by-design' instead of `post-hoc'.







\printbibliography


\end{document}